\documentstyle[prd,aps]{revtex}

\begin{document}

\draft

\title{
A priori mixed baryons and weak radiative decays
}

\author{
A.~Garc\'{\i}a\cite{email1}
}
\address{
Departamento de F\'{\i}sica.\\
Centro de Investigaci\'on y de Estudios Avanzados del IPN.\\
A.P. 14-740, C.P. 07000, M\'exico, D.F., MEXICO.\\
}
\author{
R.~Huerta\cite{email3}
and
G.~S\'anchez-Col\'on\cite{email2}
}
\address{
Departamento de F\'{\i}sica Aplicada.\\
Centro de Investigaci\'on y de Estudios Avanzados del IPN.\\
Unidad M\'erida. \\
A.P. 73, Cordemex 97310, M\'erida, Yucat\'an, MEXICO.\\
}

\date{February 6, 1996}

\maketitle

\begin{abstract}
A priori mixings of eigenstates in physical states are quantum mechanical
effects well known in several realms of physics.
The possibility that such effects are also present in particle physics, in the
form of flavor and parity mixings, is studied.
An application to weak radiative decays of hyperons is discussed.
It is suggested that this scheme may also be present in non-leptonic and rare
mode decays as the enhancement phenomenon.
\end{abstract}


\section{Introduction}
\label{introduction}

Because parity and strong flavors (strangeness, charm, etc.) are violated in
nature, the physical (mass eigenstates) hadrons cannot be either parity or
flavor eigenstates, i.e., the former must be admixtures of the latter.
It is generally believed that the breaking of flavor global groups is caused by
the mass differences of hadrons, but in such a way that parity and all flavors
are conserved, i.e., the mass operator of hadrons giving rise to such breakings
does not contain a piece that violates parity and flavor.
The flavor and parity mixings in physical hadrons are attributed to the
perturbative intervention of $W^{\pm}_{\mu}$ and $Z^0_{\mu}$ (parity mixing
only).
And, precisely because such intervention is perturbative, such mixings can
appear only in higher orders of perturbation theory; thus, such mixings appear,
so to speak, {\em a posteriori.}

However, the possibility that the mass operator of hadrons does contain a
(necessarily) very small piece that is flavor and parity violating is not
excluded by any fundamental principle.
If such a piece does exist, then, the parity and flavor admixtures in hadrons
must come {\em a priori}, in a non-perturbative way.
It is not idle to emphasize that such a piece could not be attributed to the
$W^{\pm}_{\mu}$ and $Z^0_{\mu}$.

Our purposes in this paper are (i) to explore the possibility that {\em the
mass operator of hadrons contain flavor and parity violating pieces leading to
a priori mixings,} (ii) to study how to implement the a priori mixings in
hadrons, and (iii) to illustrate the potential usefulness such mixings might
have.
Accordingly, in Sec.~\ref{ansatz} we discuss how a priori mixings may
be introduced at the hadron level via an ansatz, and in Sec.~\ref{application}
we apply a priori mixings to weak radiative decays of hyperons in order to show
how the framework we introduced can be used.
We reserve the last section to discuss the potential implications of a priori
mixings in particle physics.

To close this section, let us remark that a priori mixings are quantum
mechanical effects well known in other realms of physics, e.g., atomic physics.
Thus, another way to put the aims of this paper is to explore the questions
whether a priori mixings are also present in particle physics and what
consequences this could have.

\section{An ansatz}
\label{ansatz}

The implementation of a priori mixings for practical applications cannot, as
of today, be achieved from first principles, i.e., by starting from a model at
the quark level and then performing the QCD calculations to obtain the
physical hadrons and their couplings.
In order to proceed we must elaborate an ansatz.
We shall do this in a series of steps (or working hypothesis) and we shall
restrict what follows to spin 1/2 baryons.

Our ansatz consists of the following steps:

S1.
In addition to ordinary or $s$-baryons there exist $p$-baryons.
Let us assume that the $s$-baryons have intrinsic parity opposite to the one of
the $p$-baryons.
This is a crucial assumption in our approach.
The indeces $s$ and $p$ refer to this, $s$ means positive intrinsic parity and
$p$ means negative intrinsic parity.
Both sets have the same strong-flavor assignment and belong to two different 20
representations of $SU_4$.

S2.
There exist very small flavor and parity violating pieces in the mass operator
for such baryons and the passage to the physical baryons is performed by a
final rotation $R=(r_{ij})$ that diagonalizes the mass operator.
$R$ will be considered real for simplicity and since we are not taking into
account the $CP$-violation problem in baryon decays.
This leads to a priori flavor and parity admixtures in the physical (mass
eigenstates) baryons, for example, like
$\Lambda_{ph}=\Lambda_s+\alpha n_s+\alpha'n_p+\beta\Xi^0_s+
\beta'\Xi^0_p+\cdots$.
We do not know how to fix the matrix elements of $R$, but on experimental
grounds we can advance that the mixing angles are very small, so that,
$r_{ij}=\delta_{ij}+\epsilon_{ij}$, with $\epsilon_{ji}=-\epsilon_{ij}$
and $i,j=1,\ldots,40$.

S3.
The small mixing parameters ($\alpha$, $\alpha'$, $\beta$, etc.) are determined
by assigning strong-flavor group properties to the transformation matrix $R$.
For example, for $SU_3$ octets:

\begin{equation}
R=1 + aU_+ + bU_- + cO_{+} + dO_{-}
+ a'\hat{U}_+ + b'\hat{U}_- + c'\hat{O}_{+} + d'\hat{O}_{-}
+ \cdots,
\label{cuatrop}
\end{equation}

\noindent
where $U_{\pm}$, $\hat{U}_{\pm}$, $O_{\pm}$, and $\hat{O}_{\pm}$, are all
$U$-spin (charge conserving) ladder operators, with $U_{\pm}$ and $O_{\pm}$
($\hat{U}_{\pm}$ and $\hat{O}_{\pm}$) acting on $s$-baryons ($p$-baryons).
The $U_{\pm}$ and $\hat{U}_{\pm}$ operators connect hadrons in the same
representation, so that, they are generators, but $O_{\pm}$ and
$\hat{O}_{\pm}$ are not, of necessity, because they can  connect hadrons
that belong to different representations.
With the property $RR^{\dagger}=R^{\dagger}R=I$ and if we choose the symmetric
$D$-type couplings of $O_{\pm}$ and $\hat{O}_{\pm}$ equal to zero, then
the a priori flavor and parity mixings for $SU_3$ octets can be described
in terms of only four independent mixing angles named: $\sigma$, $\delta$,
$\delta'$, and $\hat{\sigma}$.
We must  point out that the previous rules in this step have a parallelism
at the quark level so that they should be necessary to develop a formulation
at that level.
This matter will not be tried here.

Step S3 leads to~\cite{cuatro}

\[
p_{ph} =
p_s +
\sigma\Sigma^+_s +
\delta\Sigma^+_p
+ \cdots
\]

\[
\Sigma^+_{ph} =
\Sigma^+_s -
\sigma p_s +
\delta' p_p
+ \cdots
\]

\[
\Sigma^-_{ph} =
\Sigma^-_s +
\sigma\Xi^-_s +
\delta\Xi^-_p
+ \cdots
\]

\begin{equation}
\Xi^-_{ph} =
\Xi^-_s -
\sigma\Sigma^-_s +
\delta'\Sigma^-_p
+ \cdots
\label{cinco}
\end{equation}

\[
n_{ph} =
n_s +
\sigma(\frac{1}{\sqrt 2}\Sigma^0_s +
            \sqrt{\frac{3}{2}}\Lambda_s) +
\delta(\frac{1}{\sqrt 2}\Sigma^0_p +
            \sqrt{\frac{3}{2}}\Lambda_p)
+ \cdots
\]

\[
\Lambda_{ph} =
\Lambda_s +
\sigma\sqrt{\frac{3}{2}}(\Xi^0_s - n_s) +
\delta\sqrt{\frac{3}{2}}\Xi^0_p +
\delta'\sqrt{\frac{3}{2}}n_p
+ \cdots
\]

\[
\Sigma^0_{ph} =
\Sigma^0_s +
\sigma\frac{1}{\sqrt 2}(\Xi^0_s - n_s) +
\delta\frac{1}{\sqrt 2}\Xi^0_p +
\delta'\frac{1}{\sqrt 2}n_p
+ \cdots
\]

\[
\Xi^0_{ph} =
\Xi^0_s -
\sigma
(\frac{1}{\sqrt 2}\Sigma^0_s + \sqrt{\frac{3}{2}}\Lambda_s) +
\delta'
(\frac{1}{\sqrt 2}\Sigma^0_p + \sqrt{\frac{3}{2}}\Lambda_p)
+ \cdots
\]

\noindent
We have displayed only the predominantly ordinary matter physical baryons in
terms of baryons that correspond to $SU_3$ octets, so that only three
independent mixing angles $\sigma$, $\delta$, and $\delta'$ survive in this
calculation.
The mixings with the other baryons corresponding to the 20 representations of
$SU_4$ are similar to the above ones.
In Eqs.~(\ref{cinco}) the dots stand for the latter flavor and parity mixings.

We have in mind an application to the observed weak radiative decays of
hyperons.
In this respect we introduce two more steps.

S4.
The e.m.\ current operator $J^{em}_{\mu}$ for baryons is a flavor conserving
Lorentz proper vector.

S5.
The leading form factors $f_1$ in the matrix elements of $J^{em}_{\mu}$ between
$s$ and $s$, $s$ and $p$, and $p$ and $p$ baryons are governed by the
e.m.\ charge ope\-rator and the induced form factors $f_2$ are independent of
the $s$ and $p$ indeces (because of hermiticity, the sign of $f_2$ in the
matrix elements between $p$ and $s$ baryons must be reversed w.r.t.\ the sign
of $f_2$ in the matrix elements between $s$ and $p$ baryons).

We wish to caution the reader that in assumption $S$5 the subindices $s$ and
$p$ in the form factors $f_2$ should not be confused and taken to mean that
they correspond to transition matrix elements between predominantly ordinary
matter baryons and predominantly mirror matter baryons.
This is important because the dimensionful magnetic-type $f_2$ depend on a mass
scale determined by the masses of the physical baryons used.
In Eqs.~(\ref{cinco}) the masses are of the order of 1 GeV and the pieces of
the matrix elements of $J^{em}_\mu$ between these baryons that carry the
indeces $s$ and $p$ have a mass scale of this 1 GeV order.
If one were to compute transitions between a predominantly ordinary matter
baryon and a predominantly  mirror matter baryon then, of course, the mass
scale would be dominated by the mass of the latter baryon, a scale which is
unknown and by necessity  must be very large.
In the next section we shall be concerned with transitions between
predominantly ordinary matter baryons exclusively.

\section{An application}
\label{application}

Our paper would not be complete if we did not attempt an application of the
physical baryons with the non-perturbative a priori mixings of flavor and
parity eigenstates.
A most direct application we may have is the weak radiative decays of
hyperons, although admittedly these may not necessarily be the easiest
physical processes to understand.

The important point to remark is that, in contrast to $W^{\pm}_{\mu}$ mediated
weak radiative decays, a priori mixed baryons can produce weak radiative
decays via the ordinary electromagnetic interaction hamiltonian
$H^{em}_{int}=eJ^{em}_{\mu}A^{\mu}$,
where $J^{em}_{\mu}$ is the familiar e.m. current operator which is a flavor
conserving Lorentz proper four-vector.
That is, a priori mixings in baryons lead to weak radiative decays that in
reality are ordinary parity and flavor conserving radiative decays, whose
transition amplitudes are non-zero only because physical baryons are not
flavor and parity eigenstates.
Nevertheless, we use the standard notation \lq\lq weak radiative decays" to
bring the attention of the experts in this area.

The radiative decay amplitudes we want are given by the usual matrix elements
$\langle\gamma,B_{ph}|H^{em}_{int}|A_{ph}\rangle$, where
$A_{ph}$ and $B_{ph}$ stand for hyperons.
A very simple calculation leads to the following hadronic matrix elements

\[
\langle p_{ph} | J^{\mu}_{em} | \Sigma^+_{ph} \rangle =
\bar u_p [ \sigma ( f^{\Sigma^+}_2 - f^p_2 ) +
           (\delta'f^p_2 - \delta f^{\Sigma^+}_2) \gamma^5 ]
i\sigma^{\mu\nu}q_{\nu} u_{\Sigma^+}
\]

\[
\langle \Sigma^-_{ph} | J^{\mu}_{em} | \Xi^-_{ph} \rangle =
\bar u_{\Sigma^-} [ \sigma ( f^{\Xi^-}_2 - f^{\Sigma^-}_2 ) +
                    (\delta' f^{\Sigma^-}_2 - \delta f^{\Xi^-}_2) \gamma^5 ]
i\sigma^{\mu\nu}q_{\nu} u_{\Xi^-}
\]

\begin{eqnarray}
\langle n_{ph} | J^{\mu}_{em} | \Lambda_{ph} \rangle
& = &
\bar u_n
\left\{
\sigma \left[ \sqrt{\frac{3}{2}} ( f^{\Lambda}_2 - f^n_2 ) +
                   \frac{1}{\sqrt 2} f^{\Sigma^0\Lambda}_2 \right]
\right.
\nonumber \\
& &
\left.
+ \left[
\sqrt{\frac{3}{2}}
(\delta' f^n_2  - \delta f^{\Lambda}_2)
-
\delta \frac{1}{\sqrt 2} f^{\Sigma^0\Lambda}_2
\right]
\gamma^5
\right\}
i\sigma^{\mu\nu}q_{\nu} u_{\Lambda}
\label{seis}
\end{eqnarray}

\begin{eqnarray}
\langle \Lambda_{ph} | J^{\mu}_{em} | \Xi^0_{ph} \rangle
& = &
\bar u_{\Lambda}
\left\{
\sigma \left[ \sqrt{\frac{3}{2}} ( f^{\Xi^0}_2 - f^{\Lambda}_2 ) -
         \frac{1}{\sqrt 2} f^{\Sigma^0\Lambda}_2 \right]
\right.
\nonumber \\
& &
\left.
+ \left[
\sqrt{\frac{3}{2}} (\delta' f^{\Lambda}_2 - \delta f^{\Xi^0}_2 )
+
\delta' \frac{1}{\sqrt 2} f^{\Sigma^0\Lambda}_2
\right]
\gamma^5
\right\}
i\sigma^{\mu\nu}q_{\nu} u_{\Xi^0}
\nonumber
\end{eqnarray}

\begin{eqnarray}
\langle \Sigma^0_{ph} | J^{\mu}_{em} | \Xi^0_{ph} \rangle
& = &
\bar u_{\Sigma^0}
\left\{
\sigma \left[ \frac{1}{\sqrt 2} ( f^{\Xi^0}_2 - f^{\Sigma^0}_2 ) -
         \sqrt{\frac{3}{2}} f^{\Sigma^0\Lambda}_2 \right]
\right.
\nonumber \\
& &
\left.
+ \left[
\frac{1}{\sqrt 2} (\delta' f^{\Sigma^0}_2 - \delta f^{\Xi^0}_2 )
+
\delta' \sqrt{\frac{3}{2}}f^{\Sigma^0\Lambda}_2
\right]
\gamma^5
\right\}
i\sigma^{\mu\nu}q_{\nu} u_{\Xi^0}
\nonumber
\end{eqnarray}

\noindent
The spinors $u_A$, $A=p,\Sigma^+$, etc. are ordinary four-component Dirac
spinors and $q=p_B-p_A$.
In accordance with S5, in Eqs.~(\ref{seis}) we have used the generator
properties of the electric charge, which require
$f^p_{1s}=f^{\Sigma^+}_{1s}=1$, etc. and also, since $s$ and $p$ baryons
belong to different irreducible representations,
$f^p_{1sp}=f^{\Sigma^+}_{1sp}=0$, etc.
In addition, we have dropped the indices $s$ and $p$ in the $f_2$, so that
$f^p_{2s}=f^p_{2sp}\ne f^{\Sigma^+}_{2s}=f^{\Sigma^+}_{2sp}$, etc.
All the matrix elements are of the form
$\bar u_B (C + D\gamma^5)i\sigma^{\mu\nu}q_{\nu} u_A$,
where $C$ and $D$ would, respectively, correspond to the parity conserving and
parity violating amplitudes of the $W^{\pm}_{\mu}$ mediated decays, although
in our case both amplitudes are indeed parity conserving.
Notice that Eqs.~(\ref{seis}) comply with e.m.\ gauge invariance.

We shall compare Eqs.~(\ref{seis}) with experiment, ignoring the contributions
of $W^{\pm}_{\mu}$ amplitudes.
We shall do this in order to be able to appreciate to what extent a priori
mixings provide on their own right a framework to describe weak radiative
decays.

To be able to proceed, we must decide what are the $f_2$ form factors in
Eqs.~(\ref{seis}).
They are anomalous magnetic moment transition form factors, because, for
example, $f^{\Sigma^+}_2$ corresponds to a form factor between $\Sigma^+$
flavor eigenstates present in the incoming physical $\Sigma^+$ with mass
$m_{\Sigma^+}$ and in the outgoing physical $p$ with mass $m_p$.
The $f_2$ form factors are affected by the masses of physical states.
However, we shall assume that as a first approximation such mass dependence
may be ignored.
In this case, the $f_2$ in Eqs.~(\ref{seis}) may be identified with the
measured anomalous magnetic moments of the hyperons, i.e.,
$f^A_2=\mu^{exp}_A-e_A/e_p$ (in nuclear magnetons).
Only $f^{\Sigma^0}_2$ is not measured~\cite{cinco}, we shall use its
$SU_3$ estimate,
$ \mu_{\Sigma^0} = ( \mu_{\Sigma^+} + \mu_{\Sigma^-} ) / 2 $,
as its central value with a $10\%$ error bar.
Also, we allow a $6\%$ theoretical error in all the others.

The unknown quantities in Eqs.~(\ref{seis}) are $\sigma$, $\delta$, and
$\delta'$.
We have no theoretical argument available to try to fix their values.
We must leave them as free parameters and extract their values from experiment.
For this purpose amplitudes~(\ref{seis}) should be plugged into the usual
formulas for the decay rates and angular asymmetries.
These formulas and the experimental data can be found in Ref.~\cite{cinco}.
The results are displayed in Table~\ref{table1}.
The values obtained for the a priori mixing angles are

\begin{eqnarray}
\sigma &=& (1.4 \pm 0.3) \times 10^{-6}
\nonumber \\
\delta &=& (-0.35 \pm 0.13) \times 10^{-6}
\label{siete}\\
\delta' &=& (-0.22 \pm 0.13) \times 10^{-6}
\nonumber
\end{eqnarray}

From Table~\ref{table1} one can see that, given its simplicity, the above
weakly mixed baryon scheme provides a qualitative reasonable description of
weak radiative decays of hyperons.
For completeness, our results may be compared with those obtained when the
$W$-boson is responsible for these decays.
This path has been extensively discussed, very recent reviews are found in
Ref.~\cite{revision}.
All the models considered so far contain three or more free parameters, most
of them are fixed with non-leptonic hyperon decays data.
The main conclusion of Ref.~\cite{revision} is that we still do not have a
satisfactory theoretical explanation of weak radiative decays of hyperons.
In this respect, it is important to remark that following our approach
the calculations are appreciably simpler.

Nevertheless, it must be stressed that these results must be taken only as
qualitative and not as quantitative.
Given the simplicity of the above approach we find them encouraging enough as
to take the a priori mixings in hadrons as a serious possibility.

\section{Discussion and Conclusions}
\label{discussion}

In the previous sections we have explored the possibility that flavor and
parity violating pieces in the mass operator of hadrons may exist.
In this case, physical hadrons would show non-perturbative mixings of flavor
and parity eigenstates, i.e., right from the start.
These we have called a priori mixings to distinguish them from the mixings
originated by the intervention of the $W^{\pm}_{\mu}$ and $Z^0_{\mu}$ bosons,
which are perturbative and lead to such mixings in hadrons, but in an a
posteriori fashion.

If a priori mixings are present, then weak decays may go via the flavor and
parity conserving hamiltonians of strong and electromagnetic interactions.
That is, with these mixings there would exist another mechanism to produce
weak radiative, non-leptonic, and rare mode decays of hadrons, in addition to
the already existing mechanisms provided by the $W^{\pm}_{\mu}$ and $Z^0_{\mu}$
bosons.
One is immediately led to several questions: if a priori mixings in hadrons do
exist in nature, how do their contributions compare to those of the
$W^{\pm}_{\mu}$ ?, can their contributions be relevant?, and if so, would they
improve our understanding of weak decays of hadrons?

Before discussing these questions one must first be able to calculate such
contributions.
This is not an easy task; however, one can introduce working hypotheses, based
on educated guesses as much as possible.
This we have done in Sec.~\ref{ansatz} for spin 1/2 baryons.
This collection of working hypotheses or ansatz enabled us to perform some
calculations.
As an illustration, we made an application to weak radiative decays of
hyperons, in Sec.~\ref{application}.
In order to keep things still at a relatively simple level, we introduced
some approximations and, because of this, the results obtained should be judged
as qualitative only.
We find them to be encouraging enough as to conclude that a priori mixings in
hadrons should be taken seriously, as a novel possibility in Particle Physics.

Let us retake the above questions.
As we mentioned in Sec.~\ref{application}, we lack any theoretical argument to
roughly estimate the size of the a priori mixing angles.
Clearly, it could well be the case that they are non-zero, so that this new
effect does exist in Particle Physics as it does in other realms of physics,
but they are extremely small.
This would mean that with even very precise data a priori mixings would go
undetected.
In other words, the effect might exist but it would be a theoretical curiosity,
irrelevant for practical purposes.
The next possibility would be that the mixing angles be such that they lead to
observable weak decays comparable to those mediated by $W^{\pm}_{\mu}$.
In this case, one would have to face the complicated situation of disentangling
what belongs to what in describing experimental data.
The last possibility is that the a priori mixing angles be such that they lead
to contributions appreaciably larger than the corresponding ones of
$W^{\pm}_{\mu}$.
In-as-much as a priori mixings are concerned, this is the really interesting
situation.
Their experimental predictions could then be subject to conclusive tests.
Therefore, it is this last possibility we shall concentrate upon.

In the understanding of non-leptonic, weak radiative, and rare mode decays of
hadrons a long-standing problem still remains an open challenge.
This is the enhancement phenomenon.
An impressive amount of effort has been invested in trying to demonstrate that
the strong interactions that dress the hadron weak decays mediated by
$W^{\pm}_{\mu}$ are responsible for such enhancement.
The results so far are disappointing.
It is commonly believed that the reason for this failure is our inability to
compute with QCD, but once we can calculate better this problem will be solved
favorably.
Along this line of reasoning, the situation envisaged is that the
intermediation of $W^{\pm}_{\mu}$ will saturate all measurements on flavor
changing decays of hadrons and if any other mechanism exists it will
necessarily be negligibly small, e.g., a priori mixings could not go beyond
the theoretical curiosity level we just mentioned.
However, it may happen that --- once we can calculate better with QCD and
contrary to expectations --- it is demonstrated that enhancement cannot be
produced by strong interactions.
In this situation a new mechanism would be required.

This last comment provides the means to subject a priori mixings to critical
tests.
One of these is that, if they are to be an interesting effect in hadron weak
decays, they should produce the observed enhancement phenomenon.
Another very important one is that one should expect that the a priori mixing
angles show a universality-like property, i.e., that their values appear
resonably stable in different types of weak decays.
However the judgement of how these tests and others are passed or failed will
also be limited in the near future by our inability to calculate better with
QCD.
Accordingly, one should first expect to obtain relevant qualitative results
and afterwards quantitative results based on educated guesses and simple
models as we have illustrated in Secs.~\ref{ansatz} and \ref{application}.
Clearly, it is along these lines that efforts of future research in this
subject should be addressed.
Also, the contributions of $W^{\pm}_{\mu}$ should be included at some point at
a, for consistency, small level, say, by assuming that $|\Delta  I|=1/2$
amplitudes are of the same order of magnitude as the $|\Delta I|=3/2$
amplitudes.

To close this paper and in the light of this discussion, we must stress that
our application to weak radiative decays of hyperons should be taken more than
anything else just as an exercise to learn to use a priori mixings of baryons.
A more detailed analysis of these decays should be retaken later on.
Nevertheless, for the time being we may point out that the lesson in
Sec.~\ref{application} is encouraging enough so as to take with seriousness
the possibility of the existence of this effect in Particle Physics.

\acknowledgments

The authors wish to acknowledge useful discussions with
J.~L. D\'{\i}az~Cruz, P.~Kielanowski, G.~Lopez~Castro, J.~L.~Lucio,
M.~A.~P\'erez, J.~Pestieau, M.~H.~Reno, and F.~Yndurain.
This work was partially supported by CONACYT (M\'exico).

\begin{table}
\caption{
Predictions for the asymmetries and branching fractions (in units of $10^{-3}$)
of the weak radiative decays considered,
along with the eight experimental measurements from Ref.~[2].
}
\label{table1}
\begin{tabular}
{
r@{$\rightarrow$}l
d
r@{.}l@{\,$\pm$\,}r@{.}l
d
r@{.}l@{\,$\pm$\,}r@{.}l
}
\multicolumn{2}{c}{Decay} &
$\alpha_{th}$ &
\multicolumn{4}{c}{$\alpha_{exp}$} &
\multicolumn{1}{c}{Fraction $(\Gamma_i/\Gamma)_{th}$} &
\multicolumn{4}{c}{Fraction $(\Gamma_i/\Gamma)_{exp}$}
\\
\tableline
$\Sigma^+$ & $p\gamma$ &
$-$0.75 &
$-$0 & 76 & 0 & 08 &
1.3 &
1 & 25 & 0 & 07
\\
$\Xi^-$ & $\Sigma^-\gamma$ &
0.57 &
\multicolumn{4}{c}{-----} &
0.14 &
0 & 127 & 0 & 023
\\
$\Lambda$ & $n\gamma$ &
$-$0.85 &
\multicolumn{4}{c}{-----} &
1.8 &
1 & 75 & 0 & 15
\\
$\Xi^0$ & $\Lambda\gamma$ &
$-$0.23 &
0 & 4 & 0 & 4 &
1.1 &
1 & 06 & 0 & 16
\\
$\Xi^0$ & $\Sigma^0\gamma$ &
$-$0.03 &
0 & 2 & 0 & 32 &
3.2 &
3 & 5 & 0 & 4
\\
\end{tabular}
\end{table}

\end{document}